\documentclass[a4paper]{article}

\usepackage{INTERSPEECH2022}
\usepackage{multirow}
\usepackage{graphicx}
\usepackage{cite}

\title{Residual-Guided Non-Intrusive Speech Quality Assessment}
\name{Zhe Ye, Jiahao Chen, Diqun Yan}
\address{Ningbo University, China}
\email{2111082400@nbu.edu.cn, 196003641@nbu.edu.cn, yandiqun@nbu.edu.cn}

\begin{document}

\maketitle
\begin{abstract}
  This paper proposes an approach to improve Non-Intrusive speech quality assessment(NI-SQA) based on the residuals between impaired speech and enhanced speech. The difficulty in our task is particularly lack of information, for which the corresponding reference speech is absent. We generate an enhanced speech on the impaired speech to compensate for the absence of the reference audio, then pair the information of residuals with the impaired speech. Compared to feeding the impaired speech directly into the model, residuals could bring some extra helpful information from the contrast in enhancement. The human ear is sensitive to certain noises but different to deep learning model. Causing the Mean Opinion Score(MOS) the model predicted is not enough to fit our subjective sensitive well and causes deviation. These residuals have a close relationship to reference speech and then improve the ability of the deep learning models to predict MOS. During the training phase, experimental results demonstrate that paired with residuals can quickly obtain better evaluation indicators under the same conditions. Furthermore, our final results improved 31.3 percent and 14.1 percent, respectively, in PLCC and RMSE.  
\end{abstract}

\section{Introduction}

As COVID-19 spreads across the world, speech communication systems such as online conferencing applications are becoming more and more popular in offices, studies, and other scenarios. However, in the background of people who commonly use this application, the speech quality can be significantly affected by noise, reverberation, packet loss, network jitter, and other factors. Therefore, how we can know in what situations and under what disturbances speech quality will drop dramatically is essential for the specialized departments to do targeted promotion.

Most providers have used packer loss and delay to estimate speech quality in recent years. Traditional NI-SQA estimates speech quality using objective indicators such as PESQ\cite{rix2001perceptual} or the current ITU-T Recommendation for speech quality prediction POLQA\cite{polqa}. A genuine problem is the lack of clean speech to help get objective indicators in most real scenes. Unlike existing objective speech quality indicators, Subjective measures such as the mean opinion score (MOS), obtained through the subjective Absolute Category Ratings (ACR) evaluation by ITU-T Recommendation P.808\cite{P.808}, can reflect the quality of speech. A MOS is obtained by averaging all participants’ scores over a particular condition. However, such subjective measurements are not always feasible due to the time and labor costs. We desperately need a way to predict MOS, whether by deep learning to implement or others. It is more practical for speech quality monitoring. $ConferencingSpeech 2022$ challenge first proposed to the non-intrusive objective speech quality assessment in online conferencing applications. 

Due to the powerful feature representation ability of the convolutional neural network(CNN), \cite{article1,article2,article3,article4,inproceedings,inproceedings2,hakami2017machine,falk2006single} some above methods by taking the MOS as the label, training to learn from the amount of data for speech quality assessment, Avila et al.\cite{avila2019non} proposed three neural networks for MOS and compared to PESQ and other measures, Fu et al.\cite{fu2018quality} proposed Quality-Net, a novel, end-to-end speech quality model based on bidirectional long short-term memory(BLSTM) to predict speech frame-wise. Yu et al.\cite{yu2021metricnet} proposed MetricNet, which leverages label distribution learning and joint speech reconstruction learning to improve performance significantly compared to the existing non-intrusive speech quality measurement models. MOSNet\cite{lo2019mosnet} uses CNN, BLSTM, and CNN-BLSTM as a model to predict mos. All the methods were improved in the post-input stage and achieved excellent results.

\begin{figure}[t]
  \centering
  \includegraphics[width=\linewidth]{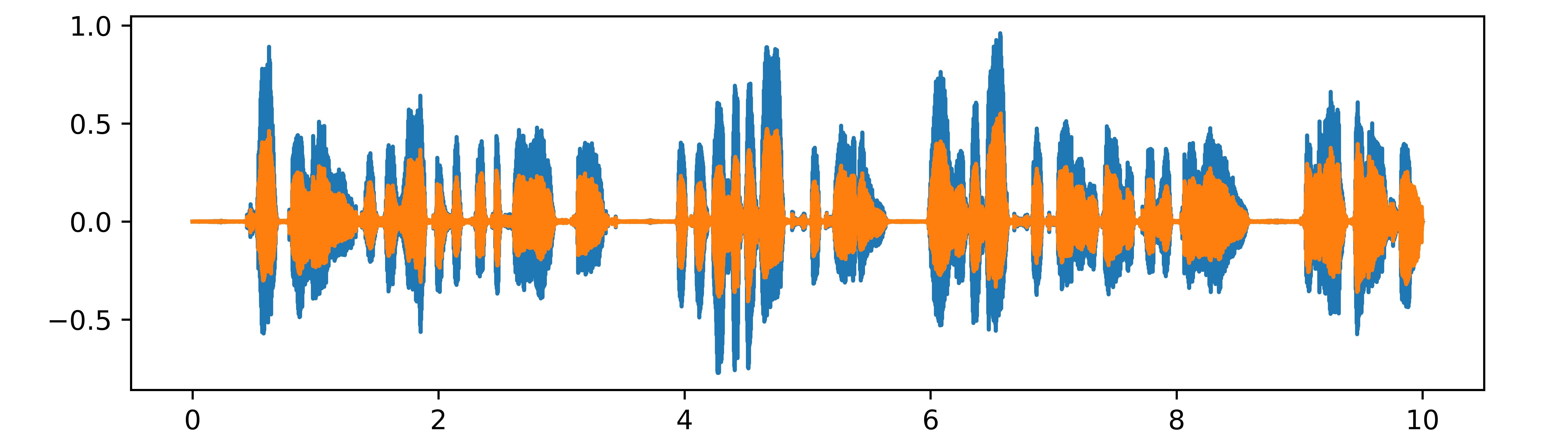}
  \caption{An illustration of impaired speech(orange) and enhanced speech(blue) using GANs.}
  \label{fig:pic1}
\end{figure}

\begin{figure*}[t]
  \centering
  \includegraphics[width=\linewidth]{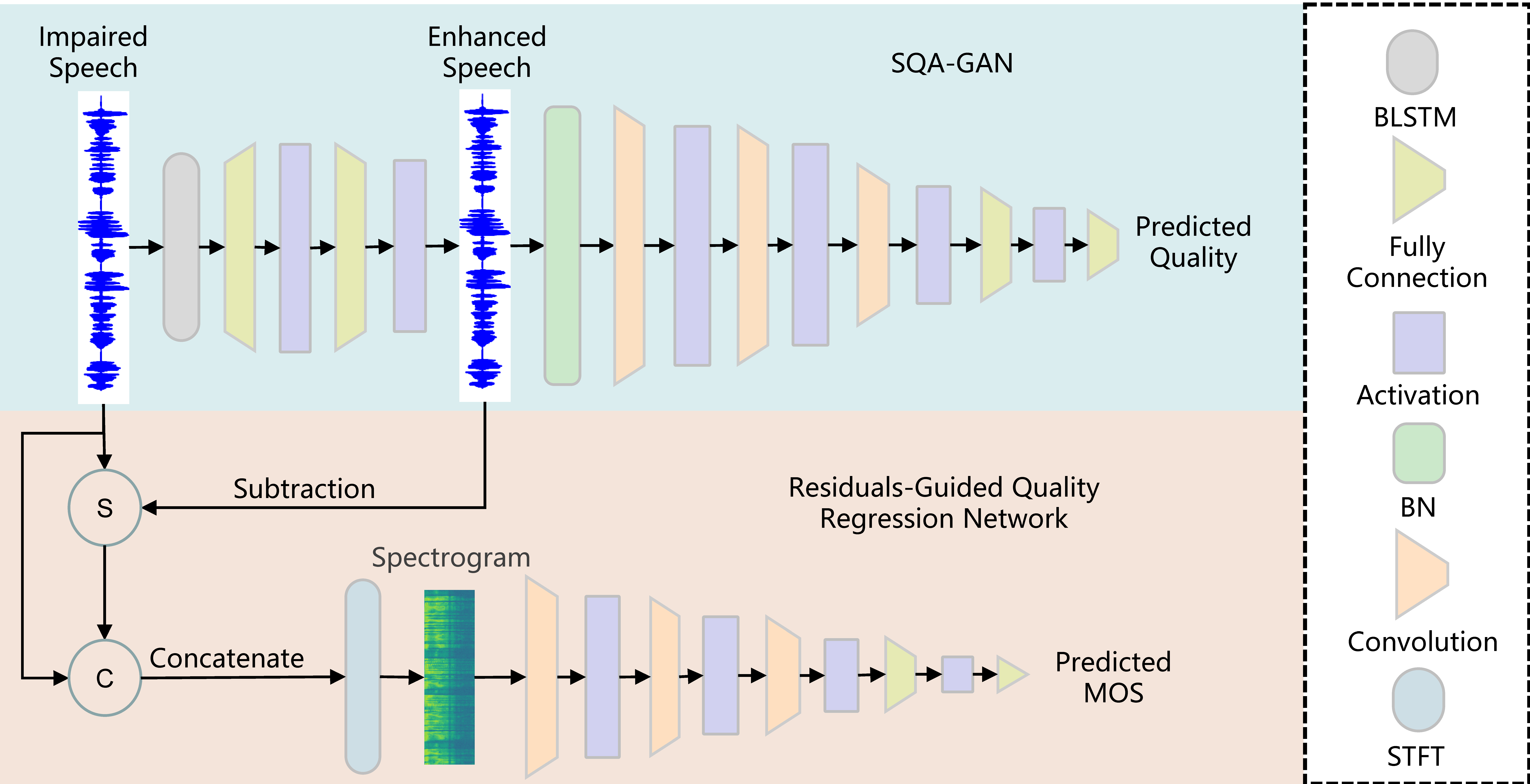}
  \caption{An illustration of our proposed Residual-Guided SQA framework.}
  \label{fig:pic2}
\end{figure*}

No reference image quality evaluation in the image domain is similar to our task, which takes only the distorted image to be assessed as input without any additional information.\cite{lin2018hallucinated,ren2018ran4iqa,ma2019no} proposed to use GAN to generate hallucinated reference, which bridges the NR-IQA and FR-IQA, allowing more methods to be applied to NR-IQA. Therefore, it’s natural to associate with the existing speech enhancement method. \cite{pascual2017segan} proposed use of generative adversarial networks for speech enhancement. \cite{fu2019metricGAN,fu2021metricgan+} proposed MetricGAN+ in which three training techniques incorporate domain knowledge of speech processing and achieve state-of-the-art results. Auditory masking refers to the effect of one sound on the auditory system's perception of another sound, which is common in nature.

The phenomenon of auditory masking plays a vital role in the perception and localization of sound in humans and animals. Still, masking effects also affect both auditory perception and auditory discrimination. Voice signals are easily polluted by noise. The communication of information by voice will be severely affected, resulting in a sharp deterioration of the performance of many communication systems. As shown in Figure \ref{fig:pic1}, speech enhancement effectively solves noise pollution and improves communication quality. The auditory masking effect is used successfully in speech enhancement, which can mask a part of the noise signal with the lower energy that enters the auditory system at the same time so that people do not perceive this part of the noise. In this paper, we proposed a novel and straightforward method, using speech residuals to improve the predicting results of MOS, and it has been preliminarily proved to be effective, especially under reverberation scenes.

Our main contributions of this work are summarised into
three folds:
\begin{itemize}
\item A novel Residual-Guided Quality Regression Network is proposed to fuse the information of residuals and impaired speeches and  help the network overcome the instability of NI-SQA and significantly improves the prediction precision and robustness.
\item A Quality-Perceiving Generative Network is proposed, in which a quality-perceiving loss is also used. We take the consideration of both sample value feature similarity and quality feature similarity in a complementary manner to generate qualified enhanced speeches. An SQA-Discriminator act as a guide to better help the learning of the generator, for the result of enhanced speeches are crucial for final prediction.
\item We evaluate the proposed method on three speech quality assessment benchmarks, including PSTN \cite{pstn}, TencentWithReverberation, and TencentWithoutReverberation. 
\end{itemize}

Our experimental results shows the superior performance of our approach among the state-of-the-art NI-SQA methods by significant margins. Comprehensive comparative experiments further demonstrates the effectiveness of our proposed method.

\section{System Description}

In this section, our approach for NI-SQA will be introduced. An overview of our framework is illustrated in Figure \ref{fig:pic2}, which consists of three parts, i.e., the quality-perceiving generative network $G$, the SQA-Discriminative network $D$ and the residual-guided quality regression network $R$. The quality-perceiving generative network $G$ produces enhanced speeches as the compensatory information for the impaired speech. The discriminative network is trained with $G$ in an adversarial manner to help $G$ produce more qualified results and constrain negative effects of bad ones to $R$. In this paper, the objective discrepancy is defined as the residuals between impaired speech and the corresponding enhanced speech. The quality regression network takes the impaired speeches and residuals as inputs, with the guidance of $G$, to exploit the perceptual discrepancy and produce the predicted quality scores as outputs to simulate the human rating about mos.

\subsection{Quality-Perceiving Generative Network}

As we mentioned above, the function of enhanced speech reference for the impaired speech is to compensate for the absence of true reference speech. The less gap between enhanced speech and true reference, the more precise the quality regression network will perform. Therefore, $G$ aims to generate a high-quality speech $S_g$ given the impaired speech $S_o$. A straightforward way to learn the generating function $G(\cdot)$ is to enforce the output of the generator both $L_2$ distance-wise and PESQ-wise(Since it is not allowed to use clean samples in $ConferencingSpeech 2022$, we did not use MOS to evaluate the generated samples.) close to the true reference. Therefore, given a set of impaired speeches $S_o$, and corresponding true reference speeches $S_t$, we solve 

\begin{equation}
\widehat{\theta}=\mathop{\arg\min}\limits_{\theta}(l_{L_2}(G_\theta(S_o),S_t)+\alpha l_p(G_\theta(S_o),S_o))
  \label{eq1}
\end{equation}
where $l_p$ penalizes the PESQ-wise differences between the output and the ground truth with PESQ-level error measurements, such as MSE, to guarantee the perceptual quality:

\begin{equation}
l_p(G_\theta(S_o),S_o)={||\phi(S_t,S_o )-\phi(G_\theta(S_o),S_o)||}^2
  \label{eq2}
\end{equation}
where $\phi(\cdot)$ represents a feature transformation. Additionally, $l_{L_2}$ penalizes the $L_2$ distance-wise differences to achieve sharper detailed results. We adopt a feature space loss term as the perception constraint, which is defined as,

\begin{equation}
l_{L_2} (G_\theta(S_o),S_t)={||G_\theta(S_o)-S_t||}^2.
  \label{eq3}
\end{equation}

\subsection{SQA-Discriminative Network}
To guarantee the generator produces high-quality outputs with less noise, especially the samples that seriously lack structure and texture information due to different reasons (e.g., decoding efficiency, latency, jitter, packet loss, or burst), we introduced the adversarial learning mechanism to our work, the original manner of which is to train $G$ to generate high-quality speeches to fool $D$, and $D$ is in contrast trained to distinguish impaired speeches $S_o$ from high-quality speeches $S_t$.

In this paper, we use $D$ to evaluate the quality of the generated speech $S_g$. However, since the impaired speeches forwarded to a quality network are usually of large size to maintain sufficient contextual information, directly providing $S_g$ as fake speeches to the discriminator will introduce instability to optimization procedure and sometimes leads to nonsensical results. More importantly, because our ultimate goal is to promote the performance of the deep regression network $R$, so even when $G$ fails to generate high-quality speeches, the predicted scores of $R$ should still be within a reasonable range. Thus, the influence of badly-enhanced speeches to $R$ should be suppressed or ignored. Therefore, we use a SQA-Discriminator to solve the problems above. As the NI-SQA models have made progress, for speech quality assessment, we use the pre-trained model like Quality-Net as our discriminator in the experiment.

\begin{table}[]
\caption{CNN design (Each conv blocks contain three convolutional layers, followed by relu and padding. The kernel size is 3x3)}
\label{tab:tab1}
\centering
\begin{tabular}{cc}
\toprule
Model                 & CNN         \\ 
\midrule
Input                 & N X 257     \\
\multirow{4}{*}{Conv} & Conv Block1 \\
                      & Conv Block2 \\
                      & Conv Block3 \\
                      & Conv Block4 \\
\multirow{4}{*}{FC}   & FC-128      \\
                      & Relu        \\
                      & Dropout      \\
                      & FC-1        \\
Output                & Pool        \\
\bottomrule
\end{tabular}
\end{table}

\subsection{Residual-Guided Quality Regression Network}
Given the enhanced speeches generated by $G$, we are able to provide references to the quality regression network to compensate for the absence of true reference information. To further exploit the information of residuals and impaired speeches effectively, which are first concatenated together, and processed into a spectrogram, the spectrogram is the final input of our residual-guided quality regression network, then we solve  

\begin{equation}
\widehat{\gamma}=\mathop{\arg\min}\limits_{\gamma} l_R(R(STFT(S_o,S_{residuals})),MOS)
  \label{eq3}
\end{equation}
where $STFT(\cdot)$ denotes short-time Fourier transform to $(S_o,S_residuals )$ as the input of the regression network $R$. Although all of the operations in Generator and regression network are differentiable and these two sub-networks can be trained in an end-to-end manner, the clean samples are not accessible as required in $ConferencingSpeech 2022$. We finally use the pretrained MetricGAN+ to enhance our speech \cite{speechbrain}. In this way, we believe that an enhanced model can amplify the effects of some unique noises. Feeding this information can significantly improve accuracy because the neural network can learn this slight difference. Then, the spectrogram will act as the input of quality regression network \cite{lo2019mosnet}, shown in Table \ref{tab:tab1}, to predict MOS. 

\begin{table*}[]
\renewcommand{\arraystretch}{1.55}
\caption{Results of overall quality in terms of RMSE, PCC, and SRCC}
  \label{tab:tab2}
  \centering
\resizebox{2.1\columnwidth}{!}{
\begin{tabular}{ccccclcccccccccc}
\toprule
\multirow{2}{*}{Dataset}   & \multicolumn{3}{c}{Baseline1} & \multicolumn{3}{c}{Baseline2}                               & \multicolumn{3}{c}{MOSNet} & \multicolumn{3}{c}{\textbf{Ours1}} & \multicolumn{3}{c}{\textbf{Ours2}}                  \\
                           & RMSE     & PCC      & SRCC    & RMSE                      & \multicolumn{1}{c}{PCC} & SRCC  & RMSE    & PCC     & SRCC   & RMSE     & PCC      & SRCC     & RMSE                      & PCC   & SRCC  \\
\midrule
Tencent w/ reverberation   & 0.467    & 0.908    & 0.889   & \multicolumn{1}{l}{0.358} & 0.890                    & 0.883 & 0.504   & 0.844   & 0.841  & 0.382    & 0.916    & 0.921    & \multicolumn{1}{l}{\textbf{0.309}}                    & \textbf{0.939} & \textbf{0.923} \\
Tencent w/o reverberation  & 0.641    & 0.885    & 0.875   & \multicolumn{1}{l}{0.415} & 0.952                   & \textbf{0.924} & 0.587   & 0.883   & 0.876  & 0.538    & 0.890    & 0.887    & \multicolumn{1}{l}{\textbf{0.382}}                     & \textbf{0.954} & 0.916 \\
PSTN Coupus                       & 0.663    & 0.770    & 0.769 & \multicolumn{1}{l}{\textbf{0.522}}            & \textbf{0.804}                   & \textbf{0.791} & 0.588   & 0.745   & 0.748  & 0.563    & 0.767    & 0.770     & \multicolumn{1}{l}{0.529} & 0.784 & 0.773\\
\bottomrule
\end{tabular}
}
\end{table*}

 \begin{figure}[t]
  \centering
  \includegraphics[width=\linewidth]{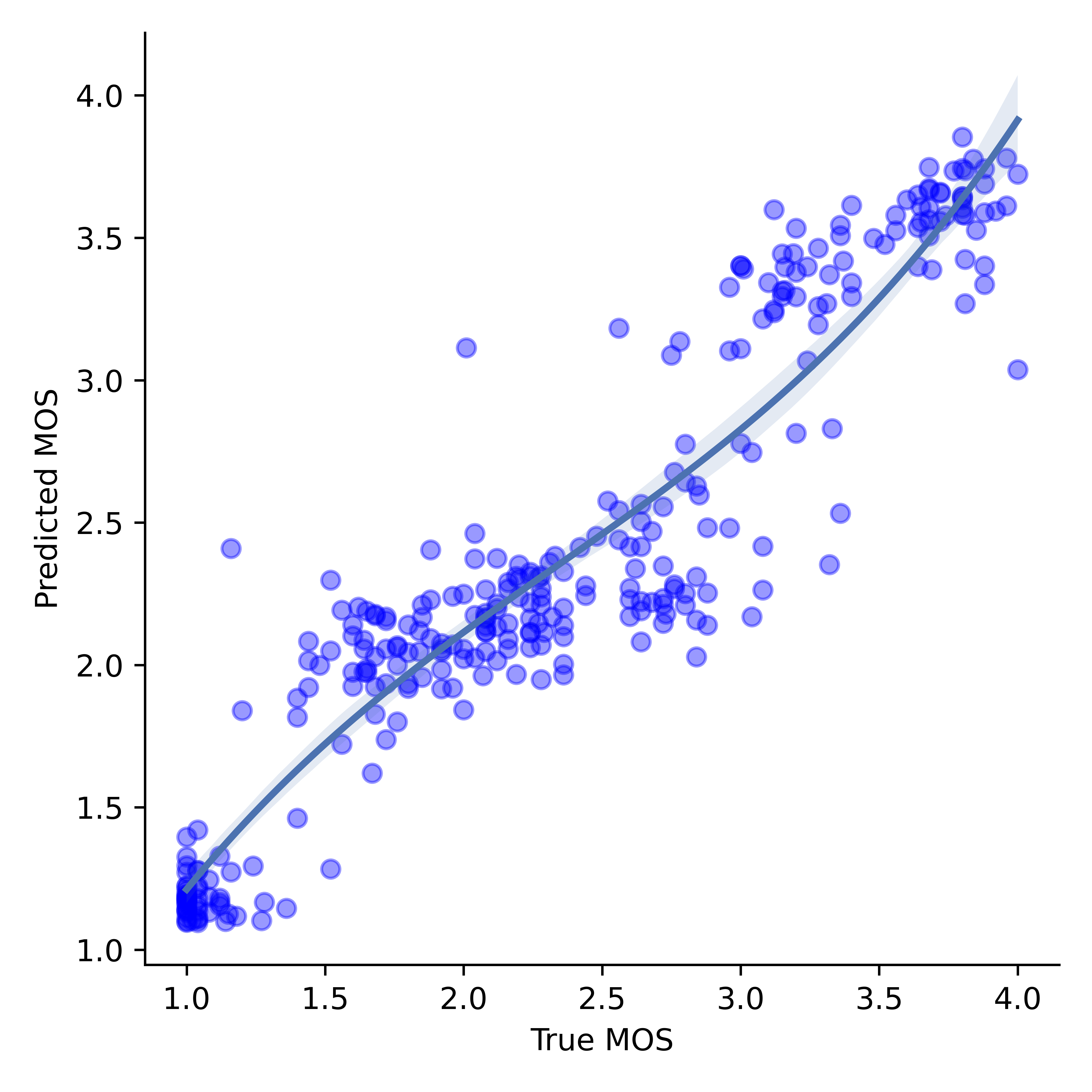}
  \caption{An illustration of true MOS and predicted MOS.}
  \label{fig:pic3}
\end{figure}

\section{Experiments}

In this section, we describe the setup of our experiments. Then, we apply our proposed method to improve different model architectures and systematically compared their results with the state-of-the-art.  Finally, based on the results, we discuss the insufficiency of our work and describe the future work to make improvements.

All models reported in this paper are trained once with the same random seed on a single NVIDIA GeForce RTX2080TI GPU and Intel Core i9-9900k clocked at 3.6GHZ.
\subsection{Experimental Setup}

\textbf{Datasets.} We perform experiments on three widely used benchmark datasets provided by $ConferencingSpeech 2022$. The entire set of speech samples (along with the corresponding MOSs) was divided into training, validation and testing, respectively. All speech samples were sampled to 16 kHz.

\textbf{Evaluation Metrics.} Following most previous works, three evaluation criteria are adopted in our paper: the Spearman’s Rank Correlation Coefficient (SRCC), the Pearson Correlation Coefficient (PCC), and the Root Mean Square Error(RMSE). SRCC measures the monotonic relationship between the MOS of ground-truth and model prediction. PCC measures the linear correlation between the MOS of the ground-truth and model prediction. RMSE measures the errors between the true MOS and predicted MOS.

\subsection{Experimental Results and Comparisons}
 We compare the results in different models. Table \ref{tab:tab2} shows the RMSE, PCC and SRCC values in different models. The  baseline1\cite{mittag2021nisqa} is adapted from the NI-SQA model and comprises a deep feedforward network followed by LSTM and average pooling. The baseline2\cite{mittag2021nisqa} trained the NI-SQA model on the entire training set is made of a CNN with self-attention network and attention-pooling. MOSNet contains three network structures, among which CNN structure was used in our experiment. For contrast, our proposed regression methods are based on CNN, the one additionaly uses the attention layer. Among all of the models, our method achieved the best RMSE, PCC on Tencent Corpus. But on PSTN Corpus, where baseline2 achieves better results than ours, our results did not behave as expected.
 
\begin{figure}[t]
  \centering
  \includegraphics[width=\linewidth]{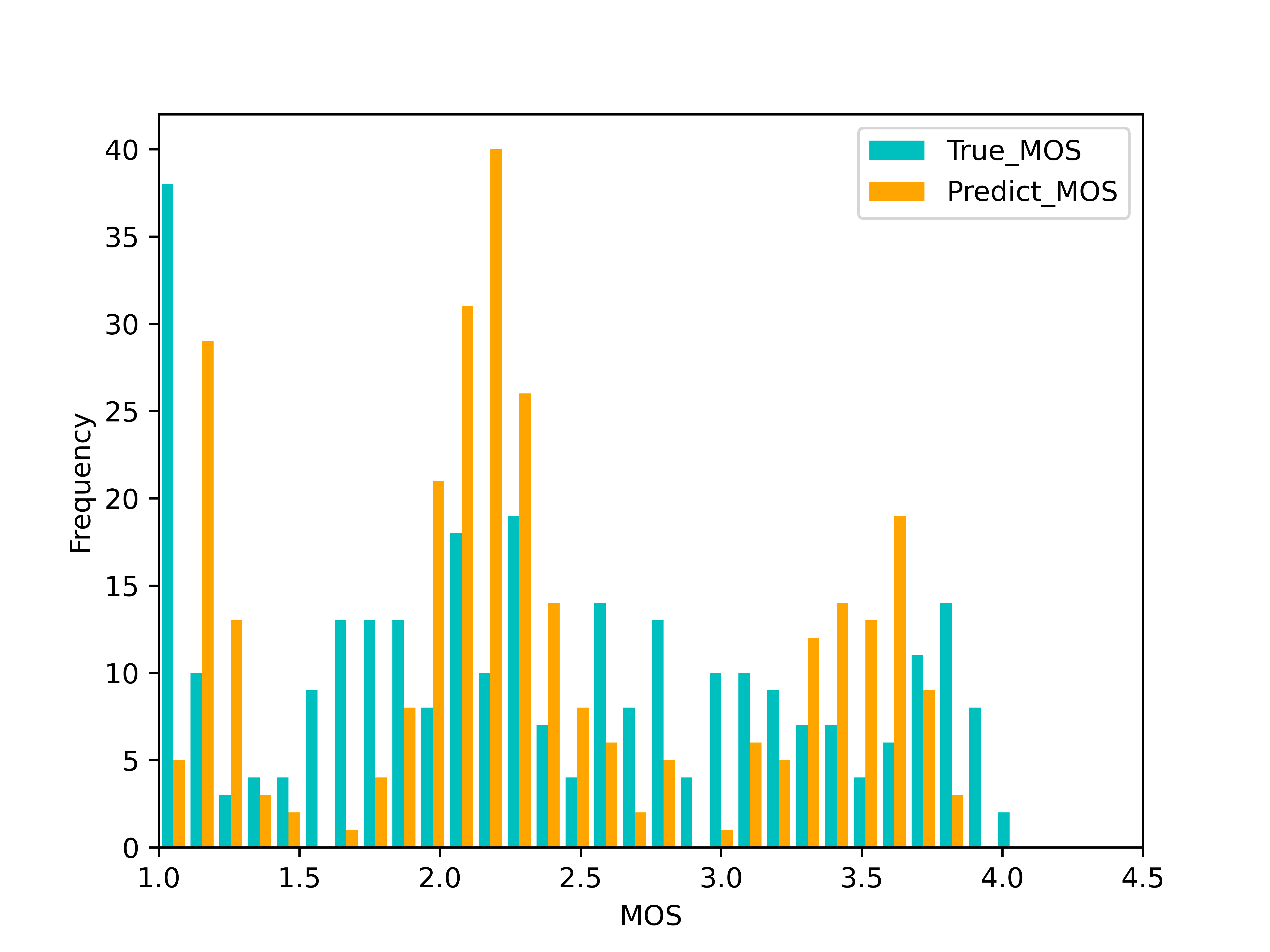}
  \caption{An illustration of the distributions of MOS.}
  \label{fig:pic4}
\end{figure}
 
 Specifically, through horizontal analysis, our method, using SQA-GAN to generate enhanced speeches to help the regression network, performs well, from the comparison between MOSNet and Ours1 and the comparison between baseline2 and ours2 on Tencent Corpus. For PSTN Corpus, Ours1 also has higher precision than MOSNet, the indicators of MOSNet have varying degrees of increasements on different datasets. Through longitudinal analysis, we can find that our method has higher stability than other state-of-the-art systems. The regression result shown in Figure \ref{fig:pic3} demonstrates that, if the error made by subjective differences in scoring is excluded, our prediction can almost take the place of human subjective opinion to score for speeches.
 
\section{Discussion}

Ideally, we should add the SQA-GAN backward into the entire neural network. Due to the limitation of competition only using the impaired speech, we apply the pre-trained GAN to enhance speeches, causing the final result may not to be the best. In future work, we will conduct further research to demonstrate our idea. Additionally, the loss function used in SQA-Discriminator can also be replaced by the evaluation of MOS, and this output can also be fused with the output of residual-guided quality regression network to get a more precise predicted value.

  Experimental results show that the indicators of our method on the Tencent dataset have been significantly improved, while the improvement on the PSTN dataset is not very obvious. After analyzing the composition of this dataset, we found that PSTN Corpus training set contains a lot of clean audio, which affects our network's learning, then indicators are not very good compared to others. Another possible reason may be that our model is too simple for PSTN Corpus to learn enough features to predict the score.
  
  In Figure \ref{fig:pic4}, we can clearly see that the data distribution is not consistent. In classification tasks, there is a Kullback–Leibler divergence, which is a method used to describe the difference between two probability distributions P and Q. Using the  Kullback–Leibler divergence as a loss, the classification network can use it to obtain as similar distributions as possible. However, there is no loss function with a similar function in the regression network.

\section{Conclusions}
In this paper, we propose a residual-guided quality regression network to solve the instability of NI-SQA models from a new perspective. We introduce SQA-GAN, an adversarial generative network trained  to generate enhanced speeches, capture the perceptual discrepancy between the impaired speeches and enhanced speeches. Then the discrepancy act as the compensate information of impaired speech to guide the quality regression network to predict precise perceptual quality results. Extra annotations or artificial prior knowledge are not require for training in our proposed network, which can be trained end-to-end. Extensive experiments demonstrate that our proposed method achieves higher correlations, lower estimation errors when compared to the other state-of-the-art systems on NI-SQA tasks. In final results of this challenge, our method reduced RMSE from 0.768 to 0.606, and improved PLCC from 0.53 to 0.696 compared to baseline.

In the follow-up research, we will use a deeper neural network and analyze the distribution of the ground-truth and predicted value to explore why the improvement is not apparent. Perhaps we can make some changes based on Kullback–Leibler divergence as a new loss function that can be used in the regression network, or we can apply the mean and variance of the true distribution of MOS to the loss function.

\clearpage
\nocite{*}
% \bibliographystyle{IEEEtran}
% \bibliography{mybib}

\end{document}